\documentclass[hyper]{JHEP}

\usepackage{latexsym}
\usepackage{amsmath}
\usepackage{amscd}
\usepackage{amsfonts}
\usepackage{amsthm}
\usepackage{graphicx}
\allowdisplaybreaks[1]

\def\IN{\mathbb{N}}

\def\IR{\mathbb{R}}

\newcommand{\tr}{\mbox{${\rm tr \;}$}}

\newsavebox{\dottedsquare}
\newsavebox{\Fcirc}
\newsavebox{\sdot}
\savebox{\Fcirc}(0,0){
\put(0,0){\circle*{1.5}}
}
\savebox{\sdot}(0,0){
\put(0,0){\circle*{0.2}}
}

\savebox{\dottedsquare}(20,20)[bl]{
\multiput(0,0)(2,0){10}{\usebox{\sdot}}
\multiput(0,0)(0,2){10}{\usebox{\sdot}}
\multiput(20,0)(0,2){11}{\usebox{\sdot}}
\multiput(0,20)(2,0){11}{\usebox{\sdot}}
\multiput(0,0)(20,0){2}{\usebox{\Fcirc}}
\multiput(0,20)(20,0){2}{\usebox{\Fcirc}}
}

\title{Testing the $\alpha'{}^3$ term in the non-abelian open superstring effective action}

\author{by Paul Koerber\thanks{Aspirant FWO} and Alexander
Sevrin\\
    Theoretische Natuurkunde, Vrije Universiteit Brussel \\
    Pleinlaan 2, B-1050 Brussels, Belgium\\
        E-mail: \email{koerber, asevrin@tena4.vub.ac.be}}
\preprint{VUB/TENA/01/07\\ \hepth{0109030}}

\abstract{Recently, a proposal for the full non-abelian open superstring
effective action through ${\cal O}(\alpha'{}^3)$ has been formulated in
{\tt hep-th/0108169}. We test this result by calculating the spectrum in
the presence of constant magnetic background fields and by comparing the
result to string theoretic predictions. The agreement is perfect. Other
proposals for the superstring effective action through this order do
not reproduce the spectrum correctly.}

\keywords{D-branes}

\begin{document}
\section{Introduction}

The tree-level effective action for a single D$p$-brane is known through all orders in
$\alpha'$, albeit in the limit of slowly varying fields. It is the $d=10$ supersymmetric
abelian Born-Infeld action, dimensionally reduced to $p+1$ dimensions \cite{pol}, \cite{BI},
\cite{susynbi}. For $n$ coinciding D$p$-branes, no all order result is known.
In leading order it is the $d=10$ $U(n)$ super Yang-Mills action, dimensionally reduced
to $p+1$ dimensions \cite{witten}.
A direct calculation requires matching the effective action to $n$-point open
superstring amplitudes. This has been done for $n\leq 4$, yielding the full effective
action through order $ {\cal \alpha}'{}^2$ \cite{direct}, \cite{direct1}, \cite{bilal}.
In the remainder of this paper, we will only focus on the bosonic terms in the action. In
addition, we will ignore the transversal coordinates of the D-brane. The latter can be
reconstructed using T-duality (see e.g. \cite{Tstr}).

In \cite{Tstr}, a detailed study of the structure 
of the non-abelian effective action was initiated. An immediate consequence was that, 
as the effective action has to match gluon disk amplitudes, there is necessarily
only one group trace. Furthermore,
it was pointed out that the effective action $ {\cal S}$ is given
by ${\cal S}={\cal S}_1+{\cal S}_2+{\cal S}_3$. In this, ${\cal S}_1$
does not contain any covariant derivatives acting on the
fieldstrength and is, by definition,
the non-abelian Born-Infeld action.
Both ${\cal S}_2$ and ${\cal S}_3$ contain the terms with derivatives acting on the
fieldstrength, but while ${\cal S}_2$ has only terms with symmetrized 
products of covariant derivatives, ${\cal S}_3$ has anti-symmetrized products
of covariant derivatives as well. It is clear that because of
the $[D,D]\cdot=[F,\cdot ]$ identity, the split between ${\cal S}_1$ and ${\cal S}_3$
is not unambiguously defined. This ambiguity was fixed in \cite{Tstr} by the proposal
that ${\cal S}_1$ is the non-abelian
Born-Infeld action defined by means of the symmetrized trace prescription.
I.e.\ it assumes the same form as the abelian Born-Infeld action. Upon expanding the action
in powers of the fieldstrength, one first symmetrizes all terms and subsequently one performs
the group trace. Indeed, all other terms without derivatives not belonging to this class can be re expressed
as elements of ${\cal S}_3$.
In the abelian limit, ${\cal S}_1$ reduces then to the abelian Born-Infeld action, ${\cal S}_3$
vanishes and ${\cal S}_2$, which is present \cite{AT},
vanishes in the limit of slowly varying fields. This proposal was
consistent with results through order $\alpha'{}^2$ where, modulo field redefinitions,
${\cal S}_2$ and ${\cal S}_3$ vanish.

In \cite{HT} and \cite{DST}, this proposal was tested. Switching on constant magnetic
background fields corresponds, upon T-dualizing, to D-branes at angles. String theory easily
allows for the calculation of the spectrum of strings stretching between different branes. In
the context of the effective action, the spectrum should be reproduced by the mass spectrum of the
off-diagonal gauge field fluctuations. In the analysis, the spectrum was calculated using ${\cal S}_1$
only. Though the spectrum was correctly reproduced through $ {\cal O}(\alpha'{}^2)$, it failed at
higher orders. This clearly shows the relevance of the
${\cal S}_2$ and ${\cal S}_3$ terms and higher order calculations are called for.
Indeed, as will become clear in this paper, such terms do contribute to the spectrum. So
contrary to the abelian case, it is hard to devise a test in the non-abelian context where derivative
terms can be ignored. 

Recently, the next order of the effective action was obtained in \cite{laatste},
where an indirect approach, developed in \cite{nieuw}, was used. A stable holomorphic bundle defines
a solution of Yang-Mills theory \cite{durham}, \cite{DUY}.
In D-brane context this corresponds to BPS
configurations of D-branes in the limit of weak fieldstrengths. Requiring
that such solutions continue to exist for arbitrary values of the fieldstrength
determines the deformation of the Yang-Mills action. Through $ {\cal O}(\alpha'{}^2)$
this yields a unique\footnote{From now on, all statements we make hold modulo
field redefinitions.} result agreeing with direct calculations. At
$ {\cal O}(\alpha'{}^3)$ a one parameter family of solutions was obtained. The parameter
could be fixed by comparing to a direct calculation of the derivative terms starting from
a four point open superstring amplitude \cite{bilal2}.  Again this result was consistent
with the proposal of \cite{Tstr}. Indeed\footnote{The terms without derivatives at this order
which were given in \cite{laatste} can all be rewritten as terms with derivatives using
the $[D,D]\cdot=[F,\cdot ]$ identity. } the $ {\cal O}(\alpha'{}^3)$ correction
falls completely in ${\cal S}_2$ and ${\cal S}_3$.

Because of the appearance of genuine derivative terms
in the effective action, one can wonder whether
the spectrum in the presence of constant magnetic background
fields gets correctly reproduced. In the present paper
we will perform this check through $ {\cal O}(\alpha'{}^3)$ and show that
the action of \cite{laatste} correctly reproduces the spectrum.
An older direct calculation of the effective action at this order, \cite{kit},
fails to do so. Recently, based on completely different grounds, another proposal 
was made for the effective action through this order \cite{milaan}. 
Here as well, the spectrum is not
correctly reproduced. We will comment on this on the last section. 

\section{The Yang-Mills spectrum}
\label{leading}

In this section we briefly recall the spectrum of Yang-Mills on
tori. We start with the lagrangian\footnote{Our metric is ``mostly
+''. Further conventions can be found in \cite{laatste}. We use
the indices $\alpha$, $\beta$, ... for the compact (and
complexified) coordinates. The indices $\mu$, $\nu$, ... run over
both compact and non-compact directions. In addition the
lagrangian, eq.\ (\ref{lag0}), should still be multiplied by an
arbitrary coupling constant $-1/g^2$. The sign arises because we
use an anti-hermitian basis for the $u(n)$ Lie algebra.},
\begin{eqnarray}
{\cal L}_{(0)}=\frac{1}{4}\tr
F_{\mu_1}{}^{\mu_2}F_{\mu_2}{}^{\mu_1}.\label{lag0}
\end{eqnarray}
We compactify $2m$ dimensions on a torus
and introduce complex coordinates, $z^\alpha=(x^{2\alpha-1}+i x^{2\alpha})/\sqrt{2}$,
$z^{\bar\alpha}=(z^\alpha)^*$, $\alpha\in\{1,\cdots,m\}$ in the compact
directions. We restrict ourselves to a $U(2)$ gauge group and switch
on constant magnetic background fields in the compact directions $ {\cal F}_{\alpha\beta}=
{\cal F}_{\bar\alpha\bar\beta}=0$, ${\cal F}_{\alpha\bar\beta}=0$ for $\alpha\neq\beta$.
So only $ {\cal F}_{\alpha\bar\alpha}$ does not vanish and we take them in the Cartan subalgebra
of $su(2)$,
\begin{eqnarray}
{\cal F}_{\alpha\bar\alpha}=
i\left(
   \begin{array}{cc}
     f_\alpha & 0 \\
     0 & -f_\alpha
   \end{array}
  \right),
\end{eqnarray}
where the $f_\alpha$, $\alpha\in\{1,\cdots,m\}$ are imaginary
constants. We write the gauge potential as, $A_\alpha={\cal
A}_\alpha+ \delta A_\alpha$, where $ {\cal A}_\alpha$ denotes the
background and  $\delta A_\alpha$ the fluctuation around it. We
are only interested in the off-diagonal fluctuations as the
diagonal ones probe the abelian part of the effective action. So
we use the following notation,
\begin{eqnarray}
\delta A=
i\left(
   \begin{array}{cc}
     0 & \delta A \\
     \delta \bar A & 0
   \end{array}
  \right),\qquad
{\cal A}=
i\left(
   \begin{array}{cc}
     {\cal A} & 0 \\
     0 & -{\cal A}
   \end{array}
  \right).
\end{eqnarray}
In this paper, we will study the fluctuations in the compact directions.
In other words only
$\delta A_\gamma$ and $\delta A_{\bar \gamma}$ are non-vanishing.
Expanding eq.\ (\ref{lag0}) through second order in the fluctuations, we find modulo a constant
term,
\begin{eqnarray}
{\cal L}^{quad}&=&-\delta\bar A_{\bar\alpha}\left({\cal D}^2+4if_\alpha
- \sum_\beta{\cal D}_\alpha( {\cal D}_\beta A_{\bar\beta}+ {\cal D}_{\bar\beta} A_{\beta})
\right)\delta A_{\alpha}- \nonumber\\
&&\delta \bar A_{\alpha}\left({\cal D}^2-4if_\alpha
- \sum_\beta{\cal D}_\alpha( {\cal D}_\beta A_{\bar\beta}+ {\cal D}_{\bar\beta} A_{\beta})
\right)\delta A_{\bar\alpha},\label{42}
\end{eqnarray}
where ${\cal D}^2={\cal D}_\mu{\cal D}^\mu$ and
\begin{eqnarray}
{\cal D}_\mu\delta A_\nu=(\partial_\mu+2i {\cal A}_\mu)\delta A_\nu, \qquad
{\cal D}_\mu\delta \bar A_\nu=(\partial_\mu-2i {\cal A}_\mu)\delta \bar A_\nu.
\end{eqnarray}
By a gauge choice, we can put
$\sum_\beta({\cal D}_\beta A_{\bar\beta}+ {\cal D}_{\bar\beta} A_{\beta})=0$.
We rewrite ${\cal D}^2\delta A_\alpha$ in eq.\ (\ref{42}), using $[ {\cal D}_{\bar\alpha}, {\cal
D}_{\beta}]=-2i\delta_{\alpha,\beta}f_\alpha$ as,
\begin{eqnarray}
{\cal D}^2\delta A_\alpha=\big(\Box_{NC}+2\sum_\beta{\cal D}_\beta
{\cal D}_{\bar\beta}-2i\sum_\beta f_\beta\big)\delta A_\alpha,\label{43}
\end{eqnarray}
where $ \Box_{NC}$ is the d'Alambertian in the non-compact directions.
So we get from eqs. (\ref{42}) and (\ref{43}),
\begin{eqnarray}
0&=&\Big(\Box_{NC}+2\sum_\beta ({\cal D}_\beta {\cal D}_{\bar\beta}-if_\beta)+4if_\alpha
\Big)\delta A_\alpha^{\{n\}}(x), \nonumber\\
0&=&\Big(\Box_{NC}+2\sum_\beta ({\cal D}_\beta {\cal D}_{\bar\beta}-if_\beta)-4if_\alpha
\Big)\delta A_{\bar\alpha}^{\{n\}}(x).\label{eom}
\end{eqnarray}
It is clear that in order to obtain
the spectrum, we need to diagonalize $\sum_\beta {\cal D}_\beta
{\cal D}_{\bar\beta}$. This was performed in \cite{spec1} and \cite{spec2}.
A complete set of eigenfunctions, $f(z,\bar z;\{n\})$, is given by
\begin{eqnarray}
f(z,\bar z;\{n_1,n_2,\cdots,n_m\})= {\cal D}_{z^1}^{n_1}{\cal D}_{z^2}^{n_2}
\cdots{\cal D}_{z^m}^{n_m}g(z,\bar z),
\end{eqnarray}
with $n_1,\,n_2,\cdots,n_m\in\IN$ and
\begin{eqnarray}
g(z,\bar z)=e^{-2i\sum_\gamma
z^{\bar\gamma} {\cal A}_{\bar\gamma}(z)} h(z).
\end{eqnarray}
The holomorphic function $h(z)$ should obey certain periodicity
conditions and was explicitly constructed in terms of
$\theta$-functions in \cite{spec2} and \cite{spec1}. As $ {\cal
D}_{\bar \gamma}g(z,\bar z)=0$, $\forall \gamma$, one easily
obtains the eigenvalues. We write
\begin{eqnarray}
\delta A_\mu(x,z,\bar z)=\sum_{\{n\}\in\IN^m}\delta A_\mu^{\{n\}}(x)
f(z,\bar z ;\{n\}),
\end{eqnarray}
where $x$ denotes the non-compact coordinates and $\{n\}=\{n_1,n_2,\cdots,n_m\}$.
From this we immediately get the dispersion relation and as a consequence the
spectrum,
\begin{eqnarray}
M^2&=& 2i \sum_\beta (2n_\beta+1 )f_\beta\pm 4 i f_\alpha .
\end{eqnarray}
String theory yields a spectrum of exactly the same form but the field strength gets
replaced everywhere by \cite{chiara} (see also \cite{DST}),
\begin{eqnarray}
f_\gamma\rightarrow \frac{1}{2\pi\alpha'}\mbox{arctanh}(2\pi\alpha'f_\gamma).
\label{strspec}
\end{eqnarray}
So for $2\pi\alpha'f_\gamma$ small, we get agreement. The higher order corrections
to the effective action should be such that the Yang-Mills spectrum gets deformed
as in eq.\ (\ref{strspec}).

\section{The spectrum through ${\cal O}(\alpha'{}^2)$}
At order $\alpha'$ only one term appears\footnote{In order to simplify the notation, we will put
after this $2\pi\alpha'=1$},
\begin{eqnarray}
{\cal L}_{(1)}=
2\pi\alpha'\xi
\tr\left(\left(D_{\mu_3}D^{\mu_1}F_{\mu_1}{}^{\mu_2}\right)
F_{\mu_2}{}^{\mu_3}\right),\label{lag1}
\end{eqnarray}
with $\xi\in\IR$. This term can be removed by a field redefinition,
\begin{eqnarray}
A_\mu\longrightarrow A_\mu - \xi D^\nu F_{\nu\mu}.
\end{eqnarray}
However, keeping this term and performing the calculation
as above, one finds the same result for the spectrum,
i.e. one recovers eq. (\ref{eom}), provided one
redefines the fluctuations as
\begin{eqnarray}
\delta A_\mu\rightarrow \delta A_\mu-2\xi \Big( {\cal D}^2\delta A_\mu
+2 [ {\cal F}_{\mu\nu}, \delta A^\nu]\Big)
\end{eqnarray}
Such redefinitions are unambiguously determined as the dispersion relation
should be of the form $(\Box_{NC}-M^2)\delta A=0$. Essential in this is the form of the
derivatives with respect to the non-compact coordinates. From now on, we will completely
ignore terms which are removable through a field redefinition.

We now turn to the $\alpha'{}^2$ contribution to the effective
action. It reads as\footnote{ The unconventional $-$ sign is due
to the fact that we chose an anti-hermitian basis for the $u(n)$
generators.},
\begin{eqnarray}
{\cal L}_{(2)}&=&-\tr\left(
\frac{1}{24} F_{\mu_1}{}^{\mu_2}F_{\mu_2}{}^{\mu_3}F_{\mu_3}{}^{\mu_4}F_{\mu_4}{}^{\mu_1} +
\frac{1}{12}F_{\mu_1}{}^{\mu_2}F_{\mu_2}{}^{\mu_3}F_{\mu_4}{}^{\mu_1}F_{\mu_3}{}^{\mu_4}-
\right. \nonumber \\
&& \left.\frac{1}{48}F_{\mu_1}{}^{\mu_2}F_{\mu_2}{}^{\mu_1}F_{\mu_3}{}^{\mu_4}F_{\mu_4}{}^{\mu_3}-
\frac{1}{96}F_{\mu_1}{}^{\mu_2}F_{\mu_3}{}^{\mu_4}F_{\mu_2}{}^{\mu_1}F_{\mu_4}{}^{\mu_3}\right).
\label{lag2def}
\end{eqnarray}
Proceeding in exactly the same way as in the previous section,
i.e.\ taking the part in ${\cal L}={\cal L}_{(0)}+{\cal L}_{(2)}$
quadratic in the derivatives and varying it with respect to
$\delta A_{\bar\alpha}$, we obtain\footnote{For our purposes it is
sufficient to study the spectrum of $\delta A_{\alpha}$ as the
spectrum of $\delta A_{\bar\alpha}$ does not yield any additional
information.},
\begin{eqnarray}
0&=&\Big(\Box_{NC}+2\sum_\beta\big(1+ \frac{1}{3}f_\beta^2\big) {\cal D}_\beta {\cal D}_{\bar\beta}
-2i\sum_\beta
(f_\beta+\frac 1 3 f_\beta^3)+4i\big(f_\alpha+\frac 1 3 f_\alpha^3\big)
\Big)\delta \hat A_\alpha+ \nonumber\\
&&\sum_\beta\big(1+\frac 1 3 f_\beta^2\big){\cal D}_\alpha( {\cal D}_\beta  A_{\bar\beta}+
{\cal D}_{\bar\beta} A_{\beta})
+ {\cal O}(\alpha'{}^4),\label{sp2}
\end{eqnarray}
where
\begin{eqnarray}
\delta \hat A_\alpha=\big(1+\frac 1 3 f_\alpha^2-\frac 1 6 \sum_\beta f_\beta^2\big)\delta  A_\alpha.
\end{eqnarray}
The last line in eq.\ (\ref{sp2}) can be eliminated by making an appropriate gauge choice,
\begin{eqnarray}
\sum_\beta\big(1+\frac 1 3 f_\beta^2\big)
\big( {\cal D}_\beta  A_{\bar\beta}+ {\cal D}_{\bar\beta} A_{\beta}\big)=0.\label{gc}
\end{eqnarray}
When passing
from $\delta A$ to $\delta\hat A$,
which is needed in order to get the leading term, $\Box_{NC}\delta A_\alpha$ correctly,
we introduced terms at order $\alpha'{}^4$ which are not relevant to
this paper. In the previous section we found that the Yang-Mills spectrum
was linear in the fieldstrengths $f_\gamma$. String theory requires the same spectrum but with the
fieldstrength $f_\gamma$ replaced by $\mbox{arctanh}(f_\gamma)=f_\gamma+\frac 1 3 f_\gamma^3
+\frac 1 5 f_\gamma^5 + \cdots$.
It is clear from eq.\ (\ref{sp2}) that the effective action provides
the correct contribution to the spectrum. In fact it was shown in \cite{STT} that knowing the abelian limit
and fitting the effective action through this order such that it reproduces the correct spectrum,
completely fixes the effective action through this order.

\section{Testing the effective action at $ {\cal O}(\alpha'{}^3)$}\label{test}
Now we finally turn to the $ {\cal O}(\alpha'{}^3)$ correction. In \cite{laatste} it
was found to be ${\cal L}_{(3)}={\cal L}^{ND}_{(3)}+{\cal L}^{D}_{(3)}$, with,
\begin{eqnarray}
{\cal L}_{(3)}^{ND}&=&-\lambda\,\tr\Big(
F_{\mu_1}{}^{\mu_2}F_{\mu_2}{}^{\mu_3}F_{\mu_3}{}^{\mu_4}F_{\mu_5}{}^{\mu_1}F_{\mu_4}{}^{\mu_5}+
F_{\mu_1}{}^{\mu_2}F_{\mu_4}{}^{\mu_5}F_{\mu_2}{}^{\mu_3}F_{\mu_5}{}^{\mu_1}F_{\mu_3}{}^{\mu_4}-
\nonumber\\
&&\frac 1 2
F_{\mu_1}{}^{\mu_2}F_{\mu_2}{}^{\mu_3}F_{\mu_4}{}^{\mu_5}F_{\mu_3}{}^{\mu_1}F_{\mu_5}{}^{\mu_4}
\Big),
\label{l3finaal}
\end{eqnarray}
and
\begin{eqnarray}
{\cal L}_{(3)}^{D}&=&\lambda\,\tr\Big(
\frac 1 2
\left(D^{\mu_1}F_{\mu_2}{}^{\mu_3}\right)\left(D_{\mu_1}F_{\mu_3}{}^{\mu_4}\right)
F_{\mu_5}{}^{\mu_2}F_{\mu_4}{}^{\mu_5} + \nonumber\\
&&\frac 1 2
\left(D^{\mu_1}F_{\mu_2}{}^{\mu_3}\right)F_{\mu_5}{}^{\mu_2}\left(D_{\mu_1}
F_{\mu_3}{}^{\mu_4}\right)F_{\mu_4}{}^{\mu_5} - \nonumber\\
&&\frac 1 8
\left(D^{\mu_1}F_{\mu_2}{}^{\mu_3}\right)F_{\mu_4}{}^{\mu_5}\left(D_{\mu_1}
F_{\mu_3}{}^{\mu_2}\right)F_{\mu_5}{}^{\mu_4} +
\left(D_{\mu_5}F_{\mu_1}{}^{\mu_2}\right)F_{\mu_3}{}^{\mu_4}\left(D^{\mu_1}
F_{\mu_2}{}^{\mu_3}\right)F_{\mu_4}{}^{\mu_5}- \nonumber\\
&&F_{\mu_1}{}^{\mu_2}\left(D^{\mu_1}F_{\mu_3}{}^{\mu_4}\right)\left(D_{\mu_5}
F_{\mu_2}{}^{\mu_3}\right)F_{\mu_4}{}^{\mu_5}\Big) ,
\label{l3finaald}
\end{eqnarray}
with $\lambda\in\IR$. Matching the derivative terms to those obtained in a direct
calculation, \cite{bilal2}, we get
\begin{eqnarray}
\lambda=- \frac{2\zeta (3)}{\pi^3}.
\end{eqnarray}
In \cite{laatste}, a detailed comparison with other results in the literature,
\cite{bilal2}, \cite{kit} and \cite{milaan}, was made. When converting to the basis we choose for expressing
the action, we found that the terms with derivatives all agreed. However at the level of the terms without
derivatives no agreement exists between \cite{laatste}, \cite{kit} and \cite{milaan}. In order to check these various
results, we will keep the terms with derivatives as in eq.\ (\ref{l3finaald}), but we will
replace eq.\ (\ref{l3finaal}) by the most general term without derivatives and leave the coefficients free,
\begin{eqnarray}
{\cal L}_{(3)}^{ND}&=&\,\tr\Big(l_{0,0}
F_{\mu_1}{}^{\mu_2}F_{\mu_2}{}^{\mu_3}F_{\mu_3}{}^{\mu_4}F_{\mu_4}{}^{\mu_5}F_{\mu_5}{}^{\mu_1}+
l_{0,1}F_{\mu_1}{}^{\mu_2}F_{\mu_2}{}^{\mu_3}F_{\mu_3}{}^{\mu_4}F_{\mu_5}{}^{\mu_1}F_{\mu_4}{}^{\mu_5}+
\nonumber\\
&&l_{0,2}F_{\mu_1}{}^{\mu_2}F_{\mu_2}{}^{\mu_3}F_{\mu_5}{}^{\mu_1}F_{\mu_3}{}^{\mu_4}F_{\mu_4}{}^{\mu_5}+
l_{0,3}F_{\mu_1}{}^{\mu_2}F_{\mu_4}{}^{\mu_5}F_{\mu_2}{}^{\mu_3}F_{\mu_5}{}^{\mu_1}F_{\mu_3}{}^{\mu_4}+
\nonumber\\
&&l_{1,0}F_{\mu_1}{}^{\mu_2}F_{\mu_2}{}^{\mu_3}F_{\mu_3}{}^{\mu_1}F_{\mu_4}{}^{\mu_5}F_{\mu_5}{}^{\mu_4}+
l_{1,1}F_{\mu_1}{}^{\mu_2}F_{\mu_2}{}^{\mu_3}F_{\mu_4}{}^{\mu_5}F_{\mu_3}{}^{\mu_1}F_{\mu_5}{}^{\mu_4}
\Big),
\label{l3}
\end{eqnarray}
with $l_{0,s},\,l_{1,s}\in\IR$ and $\lambda\in\IR$. Our result corresponds to putting
\begin{eqnarray}
&&l_{0,0}=l_{0,2}=l_{1,0}=0, \nonumber\\
&&l_{0,1}=l_{0,3}=-\lambda,\quad l_{1,1}=\frac \lambda 2.\label{ours}
\end{eqnarray}
Again, we need to expand the effective action through second order in the fluctuations. A straightforward
but very tedious calculation yields,
\begin{eqnarray}
0&=&\Big(\Box_{NC}+2\big(1+ \frac{1}{3}f_\beta^2\big) {\cal D}_\beta {\cal D}_{\bar\beta}-2i
(f_\beta+\frac 1 3 f_\beta^3)+4i\big(f_\alpha+\frac 1 3 f_\alpha^3\big)
\Big)\delta \hat A_\alpha+ \nonumber\\
&&\Big( 4(4\lambda-x_B)f_\alpha^4 +2 (x_B-4\lambda)f_\alpha^2f_\beta^2
+2(x_B-4\lambda)f_\alpha f_\beta^3+
\nonumber\\
&&2(12 \lambda-x_A)f_\beta^4
-4(\lambda+x_C) f_\beta^2f_\gamma^2
\Big)\delta A_\alpha
+ \nonumber\\
&&2i(x_B-4\lambda)f_\beta^2f_\alpha {\cal D}_\beta {\cal D}_{\bar\beta}\delta A_\alpha-
\nonumber\\
&&\Big(1+\frac 1 3 f_\beta^2+i(6\lambda-x_A)f_\alpha^3+i(4\lambda-x_B)f_\alpha^2f_\beta-2i(\lambda+x_C)f_\alpha f_\gamma^2
+\nonumber\\
&&i(x_B-4\lambda)f_\alpha f_\beta^2+
i(x_A-12\lambda)f_\beta^3+ 2i(x_C+\lambda)f_\gamma^2f_\beta
\Big) {\cal D}_\alpha {\cal D}_{\beta}\delta A_{\bar\beta}-
\nonumber\\
&&\Big(1+\frac 1 3 f_\beta^2+i(6\lambda-x_A)f_\alpha^3-i(4\lambda-x_B)f_\alpha^2f_\beta
-2i(\lambda+x_C)f_\alpha f_\gamma^2
+ \nonumber\\
&&i(x_B-4\lambda)f_\alpha f_\beta^2-
i(x_A-12\lambda)f_\beta^3- 2i(x_C+\lambda)f_\gamma^2f_\beta
\Big) {\cal D}_\alpha {\cal D}_{\bar\beta}\delta A_{\beta}+
\nonumber\\
&&\lambda f_\gamma^2\big( {\cal D}_{\gamma}{\cal D}_{\bar\gamma}{\cal D}_{\alpha}+
{\cal D}_{\bar\gamma}{\cal D}_{\gamma}{\cal D}_{\alpha}+
{\cal D}_{\gamma}{\cal D}_{\alpha}{\cal D}_{\bar\gamma}+
{\cal D}_{\bar\gamma}{\cal D}_{\alpha}{\cal D}_{\gamma}
\big)\big(
{\cal D}_{\beta}\delta A_{\bar\beta}+
{\cal D}_{\bar\beta}\delta A_{\beta}
\big)+\nonumber\\
&&{\cal O}(\alpha'{}^4),\label{sp3}
\end{eqnarray}
where
\begin{eqnarray}
\delta \hat A_\alpha&=&\delta  A_\alpha +\frac 1 3 f_\alpha^2
\delta  A_\alpha-\frac 1 6  f_\beta^2
\delta  A_\alpha
-i(x_A-4\lambda)f_\alpha^3\delta  A_\alpha- \nonumber\\
&&2 i(x_C+\lambda)f_\alpha f_\beta^2\delta  A_\alpha
+4i \lambda f_\beta^3\delta  A_\alpha
-(\lambda f_\gamma^2-2\lambda f_\alpha f_\beta) {\cal D}_{\alpha} {\cal D}_{\beta}\delta A_{\bar \beta}
\nonumber\\
&&-(\lambda f_\gamma^2+2\lambda f_\alpha f_\beta) {\cal D}_{\alpha} {\cal D}_{\bar\beta}\delta A_{ \beta}
-4\lambda f_\beta^2 {\cal D}_{\beta} {\cal D}_{\bar\beta}\delta  A_\alpha+
\lambda f_\beta^2 {\cal D}^2 \delta  A_\alpha.
\end{eqnarray}
In order to not complicate our notation unnecessarily, we understand that in the two equations
above the indices $\beta$ and $\gamma$ are summed over while the index $\alpha$ is kept fixed.
The constants $x_A$, $x_B$ and $x_C$ are expressed in terms of the coupling constants
in eq.\ (\ref{l3}),
\begin{eqnarray}
x_A&=&10 l_{0,0}-2l_{0,1}+6l_{0,2}-10l_{0,3} , \nonumber\\
x_B&=&-10 l_{0,0}+6l_{0,1}+2l_{0,2}-10l_{0,3} , \nonumber\\
x_C&=&6l_{1,0}-2 l_{1,1}.
\end{eqnarray}
It is clear from this result that terms with and without derivatives communicate with each other
in a non-trivial way. We now study eq.\ (\ref{sp3}) in detail. The first line reproduces already the correct
spectrum, so the remainder should vanish. The second and third line of eq.\ (\ref{sp3}) would alter the
zero-point energy. We get that this term vanishes iff.\
\begin{eqnarray}
x_A=12\lambda,\quad x_B=4\lambda,\quad x_C=-\lambda.\label{rc1}
\end{eqnarray}
The fourth line, which would alter the oscillator energy, vanishes then as well.
Implementing eq.\ (\ref{rc1}) in the remainder of the expression, one finds that
it vanishes by virtue of the gauge choice eq.\ (\ref{gc}). So the
$ {\cal O}(\alpha'{}^3)$ corrections to the effective action, eqs.\ (\ref{l3})
and (\ref{l3finaald}) do not alter the spectrum provided eq.\ (\ref{rc1}) holds, or equivalently,
\begin{eqnarray}
l_{0,0}&=&2\lambda-l_{0,2}+2l_{0,3}, \nonumber\\
l_{0,1}&=&4\lambda-2l_{0,2}+5l_{0,3}, \nonumber\\
l_{1,1}&=&\frac \lambda 2 + 3 l_{1,0}.\label{rc}
\end{eqnarray}
Using eq.\ (\ref{ours}), one easily checks that the effective action as proposed in
\cite{laatste} indeed reproduces the correct spectrum!

\section{Conclusions}
We can be quite confident that
\begin{eqnarray}
{\cal L}&=&-\frac{1}{4g^2}
\tr\Big(
F_{\mu_1}{}^{\mu_2}F_{\mu_2}{}^{\mu_1} -
\frac{1}{24} F_{\mu_1}{}^{\mu_2}F_{\mu_2}{}^{\mu_3}F_{\mu_3}{}^{\mu_4}F_{\mu_4}{}^{\mu_1} -
\frac{1}{12}F_{\mu_1}{}^{\mu_2}F_{\mu_2}{}^{\mu_3}F_{\mu_4}{}^{\mu_1}F_{\mu_3}{}^{\mu_4}+
 \nonumber \\
&& \frac{1}{48}F_{\mu_1}{}^{\mu_2}F_{\mu_2}{}^{\mu_1}F_{\mu_3}{}^{\mu_4}F_{\mu_4}{}^{\mu_3}+
\frac{1}{96}F_{\mu_1}{}^{\mu_2}F_{\mu_3}{}^{\mu_4}F_{\mu_2}{}^{\mu_1}F_{\mu_4}{}^{\mu_3}+
\nonumber\\
&&\frac{2\zeta (3)}{\pi^2}\,\Big(
F_{\mu_1}{}^{\mu_2}F_{\mu_2}{}^{\mu_3}F_{\mu_3}{}^{\mu_4}F_{\mu_5}{}^{\mu_1}F_{\mu_4}{}^{\mu_5}+
F_{\mu_1}{}^{\mu_2}F_{\mu_4}{}^{\mu_5}F_{\mu_2}{}^{\mu_3}F_{\mu_5}{}^{\mu_1}F_{\mu_3}{}^{\mu_4}-
\nonumber\\
&&\frac 1 2
F_{\mu_1}{}^{\mu_2}F_{\mu_2}{}^{\mu_3}F_{\mu_4}{}^{\mu_5}F_{\mu_3}{}^{\mu_1}F_{\mu_5}{}^{\mu_4}
+F_{\mu_1}{}^{\mu_2}\left(D^{\mu_1}F_{\mu_3}{}^{\mu_4}\right)\left(D_{\mu_5}
F_{\mu_2}{}^{\mu_3}\right)F_{\mu_4}{}^{\mu_5}-
\nonumber\\
&&\frac 1 2
\left(D^{\mu_1}F_{\mu_2}{}^{\mu_3}\right)\left(D_{\mu_1}F_{\mu_3}{}^{\mu_4}\right)
F_{\mu_5}{}^{\mu_2}F_{\mu_4}{}^{\mu_5} -\frac 1 2
\left(D^{\mu_1}F_{\mu_2}{}^{\mu_3}\right)F_{\mu_5}{}^{\mu_2}\left(D_{\mu_1}
F_{\mu_3}{}^{\mu_4}\right)F_{\mu_4}{}^{\mu_5} +\nonumber \\
&&\frac 1 8
\left(D^{\mu_1}F_{\mu_2}{}^{\mu_3}\right)F_{\mu_4}{}^{\mu_5}\left(D_{\mu_1}
F_{\mu_3}{}^{\mu_2}\right)F_{\mu_5}{}^{\mu_4} -
\left(D_{\mu_5}F_{\mu_1}{}^{\mu_2}\right)F_{\mu_3}{}^{\mu_4}\left(D^{\mu_1}
F_{\mu_2}{}^{\mu_3}\right)F_{\mu_4}{}^{\mu_5}\Big)\Big)
, \nonumber\\\label{finaal}
\end{eqnarray}
is the full non-abelian open superstring effective action through
${\cal O}(\alpha'{}^3)$. Indeed it is almost uniquely defined by
demanding that certain BPS configurations which are known to exist
both in the weak field limit as well as in the abelian limit,
solve the equations of motion. The only redundancy left was an
arbitrary coupling constant at ${\cal O}(\alpha'{}^3)$ which we
fixed by comparing the derivative terms in eq.\ (\ref{finaal}) to
the string theoretic calculation of these terms in \cite{bilal2}.
The effective action passes an essential test. Calculating the
spectrum in the presence of constant magnetic background fields,
correctly reproduces the string theoretic result. We verified
that various readings of the direct calculation in \cite{kit}
does not pass this test.

While the result in eq.\ (\ref{finaal}) is not sufficient to make all order predictions,
it still assumes the form suggested in \cite{Tstr} as discussed in the introduction to this
paper. The proposal of \cite{Tstr} does fix $ {\cal S}_1$. However, as we showed in this paper,
even simple checks of the effective action require not only the knowledge of
${\cal S}_1$ but that of ${\cal S}_2$ and $ {\cal S}_3$ as well.
While it is highly unlikely to get a closed expression for ${\cal S}_2$, one might hope that it 
is possible for $ {\cal S}_3$. In order to get more insight in this, we are 
constructing the next order in the effective action using the method of \cite{nieuw} and \cite{laatste}.
The reader might wonder whether the method of \cite{nieuw} and \cite{laatste} at high orders 
is any less involved than a direct superstring scattering amplitude calculation. In fact,
both become very complicated. But contrary to a direct calculation, the method
of \cite{nieuw} and \cite{laatste} lends itself for a computerized implementation. 

Finally let us turn to the calculation in \cite{milaan}. There 
the terms of the same dimensions as the ones discussed in this paper ($F^5$ and $D^2F^4$),
in the one-loop effective action of $N=4$
supersymmetric Yang-Mills in four dimensions were calculated. Using the
conversion table in \cite{laatste}, makes it possible to pass from their basis for the action to ours.
The reader can verify for himself that when doing this the resulting structure does not satisfy
eq.\ (\ref{rc}). 
This is a manifestation of the fact that in general one should not expect a direct relation between
the tree-level open string effective action and the quantum super Yang-Mills effective action (for a 
more detailed discussion, we refer to \cite{buch}). In particular, already in the abelian case, it is 
not too hard to see that the $F^8$ term in the one-loop $N=4$ super Yang-Mills effective action is different
in structure, \cite{buch}, from the $F^8$ term in the Born-infeld action \cite{ft}.

\bigskip

\acknowledgments

\bigskip

We thank Jan Troost for useful discussions. In particular, we thank Arkady Tseytlin
for making numerous suggestions and remarks and for clarifying several points which were
obscure to us.
The authors are supported in part by the
FWO-Vlaanderen and in part by the European Commission RTN programme
HPRN-CT-2000-00131, in which the authors are associated to the university
of Leuven.


\newpage

\begin{thebibliography}{99}
\bibitem{pol} J. Polchinski, \prl{75}{1995}{4724}, \hepth{9510017};
J. Dai, R. Leigh and J. Polchinski, \mpla{4}{1989}{2073}.
\bibitem{BI} E.S. Fradkin and A.A. Tseytlin, \plb{163}{1985}{123};
A. Abouelsaood, C. Callan, C. Nappi and S. Yost, \npb{280}{1987}{599};
R.G. Leigh, \mpla{4}{1989}{2767}; a detailed review is given in
A.A. Tseytlin, {\em Born-Infeld action, supersymmetry and string theory}, in
{\em The Many Maces of the Superworld}, ed. M. Shifman, World Scientific
(2000), \hepth{9908105}.
\bibitem{susynbi} M.~Cederwall, A.~von Gussich, B.~E.~W.~Nilsson and A.~Westerberg,
\npb{490}{1997}{163}, \hepth{9610148};
M.~Aganagic, C.~Popescu and J.~H.~Schwarz,
\plb{393}{1997}{311}, \hepth{9610249} and
\npb{495}{1997}{99}, \hepth{9612080};
M.~Cederwall, A.~von Gussich, B.~E.~W.~Nilsson, P.~Sundell and A.~Westerberg,
\npb{490}{1997}{179}, \hepth{9611159};
E.~Bergshoeff and P. K.~Townsend, \npb{490}{1997}{145},
\hepth{9611173}.
\bibitem{witten}  E.~Witten, \npb{460}{1996}{35},
\hepth{9510135}.
\bibitem{direct} D. J. Gross and E. Witten, \npb{277}{1986}{1}.
\bibitem{direct1} A.A. Tseytlin, \npb{276}{1986}{391} and
\npb{291}{1987}{876}.
\bibitem{bilal} E. Bergshoeff, A. Bilal, M. de Roo and A. Sevrin,
\jhep{0107}{2001}{029}, \hepth{0105274}.
\bibitem{Tstr}A.A. Tseytlin, \npb{501}{1997}{41}, \hepth{9701125}.
\bibitem{AT} O. D. Andreev and A. A. Tseytlin,
\npb{311}{1988}{205}.
\bibitem{HT} A. Hashimoto and W. Taylor, \npb{503}{1997}{193}, \hepth{9703217}.
\bibitem{DST}  F. Denef, A. Sevrin and J. Troost,
\npb{581}{2000}{135}, \hepth{0002180}.
\bibitem{laatste} P. Koerber and A. Sevrin, {\em The non-abelian Born-Infeld action
through order $\alpha'{}^3$}, preprint, \hepth{0108169}.
\bibitem{nieuw} L. De Foss\'e, P. Koerber and A. Sevrin,
\npb{603}{2001}{413}, \hepth{0103015 }.
\bibitem{durham} Such solutions appeared, as far as we know, for the first time
in the physics literature in E. Corrigan, C. Devchand, D.B. Fairlie and J.
Nuyts, \npb{214}{1983}{452}.
\bibitem{DUY} K. Uhlenbeck and S.-T. Yau, {\it Comm. Pure Appl. Math.} {\bf 39}
(1986) 257 and {\it Comm. Pure Appl. Math.} {\bf 42} (1989) 703; S.K.
Donaldson, {\it Duke Math. J.} {\bf 54} (1987) 231. See also
chapter 15 in the second volume {\em Superstring Theory},
M.B. Green, J.H. Schwarz and E. Witten, Cambridge University Press (1986).
\bibitem{bilal2} A. Bilal, {\em Higher derivative corrections to the
non-abelian Born-Infeld action}, \hepth{0106062}; see also
R.R. Metsaev and A.A. Tseytlin, \npb{298}{1988}{109}.
\bibitem{kit} Y. Kitazawa, \npb{289}{1987}{599}.
\bibitem{milaan} A. Refolli, A. Santambrogio, N. Terzi and D. Zanon, {\em $F^5$
contributions to the non-abelian Born-Infeld action from a supersymmetric
Yang-Mills five-point function}, \hepth{0105277}.
\bibitem{spec1} P. van Baal, \cmp{94}{1984}{397}
and \cmp{85}{1982}{529}.
\bibitem{spec2} J. Troost, \npb{568}{2000}{180}, \hepth{9909187}.
\bibitem{chiara} A. Abouelsaood, C. Callan, C. Nappi and S. Yost,
\npb{280}{1987}{599}
\bibitem{STT} A. Sevrin, J. Troost and W. Troost,
\npb{603}{2001}{389}, \hepth{0101192}.
\bibitem{buch} I.L. Buchbinder, S.M. Kuzenko and A.A. Tseytlin,
\prd{62}{2000}{045001}, \hepth{9911221};
I.L. Buchbinder, A.Yu. Petrov and A.A. Tseytlin, to appear.
\bibitem{ft} E.S. Fradkin and A.A. Tseytlin, \npb{227}{1983}{252}.
\end{thebibliography}
\end{document}